\documentclass[prd,showpacs,showkeys,nofootinbib,floatfix,eqsecnum,
               fleqn,preprint,12pt,tightenlines]{revtex4-1} 

\usepackage{amsmath,amssymb,revsymb,graphicx,dcolumn}

\newcommand{\beq}{\begin{equation}}
\newcommand{\eeq}{\end{equation}}
\newcommand{\beqa}{\begin{eqnarray}}
\newcommand{\eeqa}{\end{eqnarray}}
\newcommand{\bsubeqs}{\begin{subequations}}
\newcommand{\esubeqs}{\end{subequations}}

\begin{document}

\begin{widetext}
\noindent Phys. Rev. D {\bf 100}, 023536 (2019) \hfill  arXiv:1903.10450 
%
\newline\vspace*{0mm}
\end{widetext}

\title{Regularized big bang singularity}
\vspace*{1mm}

\author{F.R. Klinkhamer}
\email{frans.klinkhamer@kit.edu}

\affiliation{Institute for Theoretical Physics,
Karlsruhe Institute of Technology (KIT),\\
76128 Karlsruhe, Germany\\}

\begin{abstract}
\vspace*{1mm}\noindent
Following up on earlier work on the 
regularization of the singular Schwarzschild solution,
we now apply the same procedure to the 
singular Friedmann solution.
Specifically, we are able to remove the 
divergences of the big bang singularity, at the
price of introducing a 3-dimensional spacetime defect
with a vanishing determinant of the metric.
This particular regularization also suggests 
the existence of a ``pre-big-bang'' phase.  
\vspace*{-0mm}
\end{abstract}


\maketitle

\section{Introduction}
\label{sec:Intro}

The Friedmann solution~\cite{Friedmann1922-1924,MisnerThorneWheeler2017}
of an expanding universe 
has the so-called big bang singularity with diverging 
curvature, energy density, and temperature
[the cosmic scale factor $a(t)$ 
drops to zero at cosmic time coordinate $t=t_\text{bb}=0$].  
Quantum mechanical effects may temper these
divergences; see, e.g., Refs.~\cite{Ashtekar2010,Veneziano2017}
for two general discussions from different perspectives
(loop quantum gravity and string theory, respectively).

Awaiting the definitive theory of ``quantum gravity,'' we
propose to remain within the domain of 4-dimensional
general relativity but to allow 
for degenerate metrics. In fact, a particular degenerate metric 
has already provided a ``regularization'' of the Schwarzschild
singularity~\cite{Klinkhamer2014-mpla,Klinkhamer2014-prd,KlinkhamerSorba2014,%
Guenther2017,KlinkhamerQueiruga2018,Klinkhamer2018-jpcs},
removing the divergent behavior at the
center (radial coordinate $r=0$) at the price of
introducing a spatial 2-surface ($r=b>0$)
with a 
vanishing determinant of the metric.
In the present article, we propose doing 
something similar
with the big bang singularity 
(at cosmic time coordinate $\tau=0$, in a notation that will be  
explained later),  
by the introduction of a spatial 3-surface (at $|\,\tau\,|=b/c>0$)
with a vanishing determinant of the metric.
We emphasize, right from the start, that it is the
differential structure (rather than the topology)
which plays a crucial role for the regularization
of the big bang singularity.
Incidentally, the length scale $b$ may or may not
be related to the Planck length~\cite{Klinkhamer2007}.

With a suitable cosmic time coordinate              
$T=T(\tau) \in \mathbb{R}$ and an appropriate         
\textit{Ansatz} for a degenerate metric, we obtain  
an odd solution for the cosmic scale factor
$a(T)$, which ``jumps'' over the value $a=0$ and, thereby,
avoids the big bang singularity.
This particular $T$-odd solution may be of
interest to the recent proposal for a $CPT$-symmetric  
universe~\cite{BoyleFinnTurok2018}.
There is also a $T$-even solution $a(T)$, which is strictly
positive definite and, thus, stays away from the value $a=0$.

At this moment, it may be helpful to clarify what
we mean by ``general relativity.''
Standard general relativity is simply the theory as exposed by
Einstein in his seminal 1916 article~\cite{Einstein1916}
and elaborated upon by various textbooks such as 
Refs.~\cite{MisnerThorneWheeler2017,HawkingEllis1973}. 
In practical terms, the crucial element of general relativity is 
the Einstein gravitational field equation for the metric tensor
$g_{\mu\nu}(x)$. But Einstein makes the further assumption
(in Part~B, Sec.~8 of Ref.~\cite{Einstein1916}) 
that the determinant of the metric
vanishes nowhere, $g(x)\equiv \det g_{\mu\nu}(x) \ne 0$,
and the metric is said to be nondegenerate
(see also Sec.~2.6 of Ref.~\cite{HawkingEllis1973}).
All of
this defines \emph{standard} general relativity.

Einstein's nondegeneracy assumption is certainly reasonable 
under ``normal'' circumstances (with a metric ``not too far away'' 
from the Minkowski metric) but perhaps not under ``unusual''
circumstances, such as when spacetime singularities appear.
We propose to consider, under these unusual circumstances,
metrics which obey the standard Einstein equation but
have a vanishing determinant over a
submanifold of the spacetime manifold.
(This submanifold may be considered to correspond to
a ``spacetime defect''~\cite{Klinkhamer2014-prd,KlinkhamerSorba2014,%
Guenther2017,KlinkhamerQueiruga2018,Klinkhamer2018-jpcs},
 as will be explained 
in Sec.~\ref{sec:Modified-radiation-dominated-FLRW-universe}.)
In this sense, we use an \emph{extended} version of 
general relativity by keeping the Einstein gravitational field 
equation but allowing for degenerate metrics.
General relativity with degenerate metrics has been
considered before; see, e.g., Ref.~\cite{Horowitz1991}.
Degenerate effective metrics also appear in the context  
of condensed matter physics and are perhaps accessible
by experiment~\cite{NissinenVolovik2018,Makinen-etal2019}.

\section{Standard FLRW universe}
\label{sec:Standard-radiation-dominated-FLRW-universe}

Let us, first, review the main points of the standard 
spatially flat radiation-dominated  
Friedmann--Lema\^{i}tre--Robertson--Walker (FLRW) universe.
Details and further references
can be found in, e.g., Ref.~\cite{MisnerThorneWheeler2017}.
Greek indices run over $\{0,\, 1,\, 2,\, 3\}$ 
and Latin indices over $\{1,\, 2,\, 3\}$.
Unless stated otherwise, we set $c=1$ and $\hbar=1$.

The line element 
of the standard spatially flat FLRW universe 
in comoving coordinates reads%
\bsubeqs\label{eq:stand-FLRW}
\beqa\label{eq:stand-FLRW-ds2}   
ds^{2}\,\Big|_\text{stand.\;FLRW}
&\equiv&
g_{\mu\nu}(x)\, dx^\mu\,dx^\nu \,\Big|_\text{stand.\;FLRW}
=
-dt^{2} + a^{2}(t)\;\delta_{kl}\,dx^k\,dx^l\,,
\\[2mm]
\label{eq:stand-FLRW-t-range}
t &\in& (0,\,\infty)\,,\quad 
\\[2mm]
x^k &\in& (-\infty,\,\infty)\,,
\eeqa
\esubeqs
where the restricted range of the cosmic time coordinate $t$
will be explained shortly. The real function $a(t)$ corresponds
to the cosmic scale factor. As far as the metric is concerned,
the sign of $a(t)$ is irrelevant.

With this metric and the energy-momentum tensor of
a homogeneous perfect fluid [energy density $\rho(t)$ and pressure $P(t)$], 
the Einstein equation without a
cosmological constant $\Lambda$
gives the spatially flat Friedmann 
equation~\cite{Friedmann1922-1924}
and the energy-conservation equation:
\bsubeqs\label{eq:Friedmann-equations-abc}
\beqa\label{eq:Friedmann-equation-a}
\left( \frac{1}{a(t)}\,\frac{d a(t)}{d t} \right)^{2}
&=& \frac{8\pi}{3}\,G_N\,\rho(t)\,,
\\
[2mm]
\label{eq:Friedmann-equation-b}
\frac{d}{d a} \bigg[ a^{3}\,\rho(a)\bigg]+ 3\, a^{2}\,P(a)&=&0\,,
\eeqa
to which is added the equation of state,
\beqa
\label{eq:Friedmann-equation-c}
P=P(\rho)\,.
\eeqa
\esubeqs
Consider, for definiteness, relativistic matter,
\beq\label{eq:rel-matter}
P = \frac{1}{3}\;\rho\,,
\eeq
so that \eqref{eq:Friedmann-equation-b} implies $\rho \propto 1/a^{4}$.
The resulting cosmic scale factor from \eqref{eq:Friedmann-equation-a}
is then
\beq\label{eq:standard-Friedmann-asol}
a(t)\,\Big|_\text{stand.\;FLRW}^\text{(rel-mat.\;sol.)} = \sqrt{t/t_{0}}\,.
\eeq
For the particular solution \eqref{eq:standard-Friedmann-asol},
the zero point of the cosmic time coordinate $t$ 
has been shifted, so that%
\beq
\lim_{t\to 0^{+}} a(t)\,\Big|_\text{stand.\;FLRW}^\text{(rel-mat.\;sol.)} =0 \,,
\eeq
and $t=0$ corresponds to the big bang singularity.
In addition, the cosmic scale factor \eqref{eq:standard-Friedmann-asol}
has been normalized to $1$ at a given time $t=t_{0} > 0$
for which the Hubble constant is assumed to be
positive, $H_{0} \equiv [(da/dt)/a\,]_{t=t_{0}} >0$.

The standard FLRW spacetime manifold with 
metric \eqref{eq:stand-FLRW-ds2}  
and cosmic scale factor \eqref{eq:standard-Friedmann-asol}
has the line element
\beqa\label{eq:stand-FLRW-ds2-sol-with-t}
ds^{2}\,\Big|_\text{stand.\;FLRW}^\text{(rel-mat.\;sol.)}
&=&
- dt^{2} 
+ \sqrt{t^{2}/t_{0}^{2}}\,\;\delta_{kl}\,dx^k\,dx^l\,,
\eeqa
where the metric component $\sqrt{t^{2}/t_{0}^{2}}$ can be simplified to
$t/t_{0}$, because both $t$ and $t_{0}$ are positive. The metric of this
spacetime manifold solves the Einstein equation
for a homogeneous perfect fluid of relativistic matter, but the manifold
is geodesically incomplete. Indeed, there is
a
big bang singularity at $t=0$ with
diverging curvature (as shown by, for example, 
the Kretschmann curvature scalar 
$K \equiv R^{\mu\nu\rho\sigma}\,R_{\mu\nu\rho\sigma}$)
and diverging matter energy density $\rho$ and temperature $\mathcal{T}$.
Recall that, for the special case of relativistic matter, the
Ricci curvature scalar $R \equiv g^{\mu\nu}\,R_{\mu\nu}$ vanishes
identically. Specifically, these quantities are given by the
following expressions:
\bsubeqs\label{eq:stand-FLRW-R-K-rho-Temp-sol-with-t}
\beqa
\label{eq:stand-FLRW-R-sol-with-t}
R(t)\,\Big|_\text{stand.\;FLRW}^\text{(rel-mat.\;sol.)}
&=& 
6\, \left[\left( \frac{1}{a(t)}\,\frac{d a(t)}{d t} \right)^{2}
           +
           \frac{1}{a(t)}\,\frac{d^{2} a(t)}{dt^{2}}
     \right]
=0 \,,
\\[2mm]
\label{eq:stand-FLRW-K-sol-with-t}
K(t)\,\Big|_\text{stand.\;FLRW}^\text{(rel-mat.\;sol.)}
&=& 
12\, \left[\left( \frac{1}{a(t)}\,\frac{d a(t)}{d t} \right)^{4}
           +
           \left(\frac{1}{a(t)}\,\frac{d^{2} a(t)}{dt^{2}}\right)^{2}
     \right]
\propto 1/t^{4}\,,
\\[2mm]
\label{eq:stand-FLRW-rho-sol-with-t}
\rho(t)\,\Big|_\text{stand.\;FLRW}^\text{(rel-mat.\;sol.)}
&\propto& 1/a^{4}(t) \propto 1/t^{2}\,,
\\[2mm]
\label{eq:stand-FLRW-Temp-sol-with-t}
\mathcal{T}(t)\,\Big|_\text{stand.\;FLRW}^\text{(rel-mat.\;sol.)} 
&\propto& 1/\sqrt[4]{a^{4}(t)} \propto 1/\sqrt[4]{t^{2}}\,,
\eeqa
\esubeqs
where the final expression for the temperature can
be simplified to $1/\sqrt{t}$, because $t$  is positive.
Observe that the temperature expression
$\mathcal{T}(a) \propto \big[a^{4}\big]^{-1/4}$ 
in \eqref{eq:stand-FLRW-Temp-sol-with-t}
follows directly   
from \eqref{eq:stand-FLRW-rho-sol-with-t}
and the Stefan--Boltzmann law $\rho \propto \mathcal{T}^4$  
(see also the discussion in the last paragraph of Box.~29.2 on p.~779 
of Ref.~\cite{MisnerThorneWheeler2017}).

In view of the results \eqref{eq:standard-Friedmann-asol}
and \eqref{eq:stand-FLRW-R-K-rho-Temp-sol-with-t},
it is clear that this particular solution of the Einstein equation 
is only well behaved if the range of the cosmic time coordinate $t$ 
is restricted to the open half-line $\mathbb{R}^{+}$.

\section{Modified FLRW universe}
\label{sec:Modified-radiation-dominated-FLRW-universe}

As mentioned in Sec.~\ref{sec:Intro}, it is possible
to obtain a regularized version~\cite{Klinkhamer2014-mpla}  
of the singular Schwarz\-schild
solution~\cite{MisnerThorneWheeler2017} by a simple procedure. 
The first step is to perform local surgery on Euclidean 3-space: 
the interior of a ball with radius $b$ is
removed and antipodal points on the boundary of the ball
are identified.
The resulting 3-space $\widetilde{M}_3$ is topologically nontrivial:
$\widetilde{M}_3 \simeq \mathbb{R}P^{3} - \{\text{point}\}$,
where $\mathbb{R}P^{3}$ is the 3-dimensional real-projective plane. 
A proper solution of the Einstein equation
requires suitable coordinates over $\widetilde{M}_3$ and 
an appropriate \textit{Ansatz} for the metric.

The obtained regularized Schwarzschild solution 
has a spatial 2-surface (with topology $\mathbb{R}P^{2}$)
over which the determinant of the metric vanishes.
This spatial 2-surface embedded in spacetime
may be interpreted as a (2+1)-dimensional ``defect'' of spacetime
with topology $\mathbb{R}P^{2} \times \mathbb{R}$.
At the location of the ``spacetime defect,'' the 
standard elementary-flatness condition does not
apply,
and the equivalence principle is violated
(cf. Appendix~D in Ref.~\cite{Klinkhamer2014-mpla}).
The appearance of this spacetime defect 
is, apparently, the price to pay for the absence of
the Schwarzschild  curvature singularity
(having a degenerate metric evades certain singularity theorems; 
cf. Sec.~3.1.5 in Ref.~\cite{Guenther2017}).
The technical details of this regularization procedure
can be found in Ref.~\cite{Klinkhamer2014-mpla},
with further discussion of the differential
structure in Ref.~\cite{KlinkhamerSorba2014}
and further discussion
of the physics of this particular type of spacetime defect 
in Refs.~\cite{Klinkhamer2014-prd,KlinkhamerQueiruga2018,Klinkhamer2018-jpcs}.
Some details on the mathematics 
of general relativity with degenerate metrics
appear in Chap.~3 of Ref.~\cite{Guenther2017}
(different mathematical aspects are discussed in Ref.~\cite{Horowitz1991}).

\begin{figure}[t]
\vspace*{-0mm}
\begin{center}  
\includegraphics[width=0.5\textwidth]{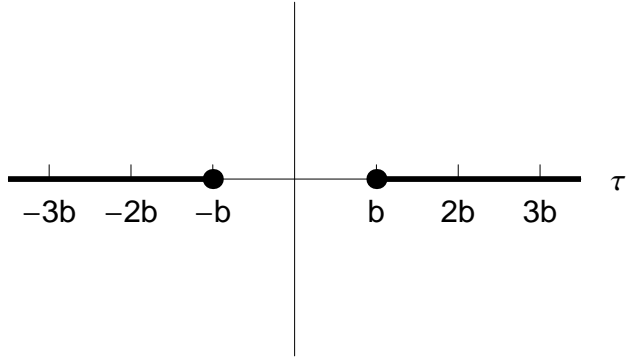}
\end{center}
\vspace*{-5mm}
\caption{Cosmic time axis 
$\tau \in  (-\infty,\,-b]  \, \cup \, [b,\,\infty)$,
where the points $\tau=-b$ and $\tau=b$ 
are identified (as indicated by the dots).}
\label{fig:sketch-ttilde-axis}
\vspace*{0mm}
\end{figure}

The idea, now, is to apply the same procedure to the singular
FLRW solution of Sec.~\ref{sec:Standard-radiation-dominated-FLRW-universe}, 
where the surgery will concern the cosmic time axis.
First, the standard cosmic time coordinate $t>0$ is replaced
by an extended coordinate $\tau \in \mathbb{R}$.
Then, surgery on this 1-space 
removes the open $\tau$ interval $(-b,\,b)$, for $b>0$,
and identifies the antipodal points $\tau=-b$ and $\tau=b$. 
In this case, the resulting 1-space $\widetilde{M}_1$
is topologically trivial:
$\widetilde{M}_1 \simeq  \mathbb{R}P^1 - \{\text{point}\}\simeq \mathbb{R}$.
A sketch is given in Fig.~\ref{fig:sketch-ttilde-axis},
which may be considered to be the 1-dimensional analog
of Fig.~1 in Ref.~\cite{Klinkhamer2014-mpla} 
for the 3-dimensional Schwarzschild construction.
Note that we will use the same symbol $b$ for the
parameter of the Schwarzschild solution and the one of
the Friedmann solution, but these length scales can, 
in principle, be different. Remember that we have set $c=1$.

Next, define a suitable cosmic time coordinate $T$ 
(not to be confused with the temperature $\mathcal{T}$ of 
 matter):
\bsubeqs\label{eq:T-def}
\beqa\label{eq:T-def-pos-neg}
T &\equiv& 
\begin{cases}
+ \sqrt[4]{\,\tau^{4} - b^{4}}\,,   &  \;\;\text{for}\;\; \tau \geq b\,,
\\[1mm]
- \sqrt[4]{\,\tau^{4} - b^{4}}\,,   &  \;\;\text{for}\;\; \tau \leq -b \,,
\end{cases}
\eeqa
\beqa
\label{eq:cosmic-time-axis-tau}
\tau &\in&  (-\infty,\,-b]  \, \cup \, [b,\,\infty)\,,
\eeqa
\esubeqs
where $b$ is assumed to be
positive.
The coordinate $T$ from \eqref{eq:T-def-pos-neg} covers the cosmic time axis
\eqref{eq:cosmic-time-axis-tau},
with a \emph{unique} value of $T$ for \emph{each} point of the axis
(cf. Fig.~\ref{fig:sketch-ttilde-axis}).
Furthermore, the expression \eqref{eq:T-def-pos-neg} 
has the same mathematical structure
as the Schwarz\-schild-construction expression (2.26) in
Ref.~\cite{KlinkhamerSorba2014}, where $y$ must be replaced by $T$
and $\pm r$ by $\tau$  and where,
following Endnote~18 of Ref.~\cite{KlinkhamerSorba2014}, 
the squares are replaced by quartic powers 
and the square roots by fourth roots.
It is, in principle, also possible 
to use $\widetilde{T} \equiv \pm \sqrt{\,\tau^{2} - b^{2}}$,
but the choice \eqref{eq:T-def-pos-neg}
has an advantage for the $T$-odd scale factor solution,
as will be explained later.

With this new cosmic time coordinate $T \in \mathbb{R}$, 
we make the following \textit{Ansatz} for the line element:%
\bsubeqs\label{eq:mod-FLRW}
\beqa\label{eq:mod-FLRW-ds2}  
ds^{2}\,\Big|_\text{mod.\;FLRW}
&=&
- \frac{T^{6}}{\big(b^{4}+T^{4}\big)^{3/2}}\;dT^{2} 
+ a^{2}(\tau)\;\delta_{kl}\,dx^k\,dx^l\,,
\eeqa\beqa
a^{2}(\tau)\,\big|_{\tau=-b}
&=&
a^{2}(\tau)\,\big|_{\tau=b}\,\,,
\eeqa\beqa
T &\in& (-\infty,\,\infty)\,, 
\eeqa\beqa
x^k &\in& (-\infty,\,\infty)\,,
\eeqa\beqa
\label{eq:mod-FLRW-tau-def}
\tau(T)&=&
\begin{cases}
+ \sqrt[4]{b^{4}+T^{4}}\,,   &  \;\;\text{for}\;\; T \geq 0\,,
 \\[1mm]
- \sqrt[4]{b^{4}+T^{4}}\,,   &  \;\;\text{for}\;\; T \leq 0\,,
\end{cases} 
\eeqa
\esubeqs
where the auxiliary coordinates $\tau=-b$ and $\tau=b$
correspond to the \emph{single} point $T=0$ on the cosmic time axis
(cf. Fig.~\ref{fig:sketch-ttilde-axis}).
The particular form of the metric \textit{Ansatz} \eqref{eq:mod-FLRW-ds2}
is inspired by the metric of the modified Schwarzschild 
solution~\cite{Klinkhamer2014-mpla}.
We notice that the coordinate transformation \eqref{eq:mod-FLRW-tau-def}  
is not a diffeomorphism, which is defined to be                           
an invertible $\text{C}^{\infty}$ function~\cite{HawkingEllis1973}.         
We also remark that, even for                                              
$a(\tau) \ne 0$, the metric from \eqref{eq:mod-FLRW-ds2} is degenerate:
$\det\,g_{\mu\nu} = 0$ at $T=0$. 
The corresponding $T=0$ spacetime slice
may be interpreted as a 3-dimensional ``defect'' of spacetime
with topology $\mathbb{R}^{3}$.
(For modified FLRW universes with positive or negative spatial curvature,
the 3-dimensional defect of spacetime has
topology $S^{3}$ or $\mathbb{H}^{3}$.)

It is straightforward to calculate the dynamic equations
by evaluating the Einstein equation without a
cosmological constant $\Lambda$
for the metric \eqref{eq:mod-FLRW-ds2} with
coordinates $\{T,\, x^1,\, x^{2},\, x^{3}\}$
and the energy-momentum tensor of
a homogeneous perfect fluid [energy density $\rho(T)$ 
and pressure $P(T)=\rho(T)/3$ for the case of relativistic matter].
But the result 
also follows from the observation that the new metric \eqref{eq:mod-FLRW-ds2}
written in terms of $\tau$ takes the same form as the
standard metric \eqref{eq:stand-FLRW-ds2} in terms of $t$.
The dynamic equations are, then, obtained if we replace $t$ with $\tau$ 
in \eqref{eq:Friedmann-equations-abc}
and change to the $T$ coordinate from \eqref{eq:T-def-pos-neg},
\bsubeqs\label{eq:mod-Friedmann-equations-abc} 
\beqa\label{eq:mod-Friedmann-equation-a}
\left(1+ \frac{b^{4}}{T^{4}}\right)^{3/2} 
\left( \frac{1}{a(T)}\,\frac{d a(T)}{d T} \right)^{2}
&=& \frac{8\pi}{3}\,G_N\,\rho(T)\,,
\\[2mm]
\label{eq:mod-Friedmann-equation-b}
\frac{d}{d a} \bigg[ a^{3}\,\rho(a)\bigg]+ 3\, a^{2}\,P(a)&=&0\,,
\\[2mm]
\label{eq:mod-Friedmann-equation-c}
P(T) &=& \frac{1}{3}\,\rho(T)\,,
\eeqa
\esubeqs
where  we have, again, considered relativistic matter.
Compared to the standard FLRW equations 
\eqref{eq:Friedmann-equation-a}, \eqref{eq:Friedmann-equation-b}, 
and \eqref{eq:rel-matter},
the only difference in \eqref{eq:mod-Friedmann-equations-abc} is 
the Jacobian factor    
$\left(d T/d \tau\right)^{2} = \left(1+ b^{4}/T^{4} \right)^{3/2}$ 
on the left-hand side of the modified Friedmann
equation \eqref{eq:mod-Friedmann-equation-a}. 
This particular prefactor in \eqref{eq:mod-Friedmann-equation-a} 
allows for a solution $a(T)$ with
$a(0)\ne 0$ and $\big[d a(T)/d T\big]^{2} \sim T^{6}$ near $T=0$.
The modified Friedmann equation \eqref{eq:mod-Friedmann-equation-a} is, 
in fact, a \emph{singular} differential equation 
(the singularity appears at $T=0$) 
with a \emph{nonsingular} solution to be given shortly,
whereas the standard Friedmann equation \eqref{eq:Friedmann-equation-a} 
is a nonsingular differential equation with a singular 
solution \eqref{eq:standard-Friedmann-asol}.  

The solutions $a(T)$ of \eqref{eq:mod-Friedmann-equations-abc} can
be even or odd in $T$. In view of the recent interest~\cite{BoyleFinnTurok2018}
in a $T$-odd solution, we explicitly give our modified
$T$-odd relativistic-matter solution $a(T)$ 
from \eqref{eq:mod-Friedmann-equations-abc},%
\beq\label{eq:mod-Friedmann-T-odd-asol}
a(T)\,\Big|_\text{mod.\;FLRW}^\text{($T$-odd\;rel-mat.\;sol.)} = 
\begin{cases}
+ \sqrt[8]{\big(b^{4}+T^{4}\big)\big/\big(b^{4}+T_{0}^{4}\big)}\,,  
& \;\;\text{for}\;\; T > 0\,,
 \\[2mm]
- \sqrt[8]{\big(b^{4}+T^{4}\big)\big/\big(b^{4}+T_{0}^{4}\big)}\,,
&  \;\;\text{for}\;\; T \leq 0\,,
\end{cases} 
\eeq
with normalization $a(T_{0})=1$ for $T_{0}>0$.
Note that the solution \eqref{eq:mod-Friedmann-T-odd-asol} 
for $T>0$ reproduces, in the limit $b\to 0^{+}$
and with the identification $T_{0}=t_{0}$, the positive-$t$ branch 
of the standard solution \eqref{eq:standard-Friedmann-asol}.

The solution \eqref{eq:mod-Friedmann-T-odd-asol}
is discontinuous at $T=0$ but
still has a monotonic behavior, $da(T)/dT \geq 0$.
Observe also that   
the $T$-odd solution \eqref{eq:mod-Friedmann-T-odd-asol}
has continuous first-, second- , and third-order derivatives at $T=0$
and a  discontinuous fourth-order derivative at $T=0$
(the derivatives must be defined appropriately).
If the $T$ definition \eqref{eq:T-def-pos-neg} were              
replaced by $\widetilde{T} \equiv \pm \sqrt{\,\tau^{2} - b^{2}}$,
the corresponding $\widetilde{T}$-odd solution $a(\widetilde{T})$ would 
already have a discontinuous second-order derivative 
at $\widetilde{T}=0$.

With a nonvanishing parameter $b$, 
the solution \eqref{eq:mod-Friedmann-T-odd-asol} 
gives finite values at $T=0$ for the Ricci curvature scalar $R$ 
(identically zero, in fact)
and the Kretschmann curvature scalar $K$,%
\bsubeqs\label{eq:Rsol-Ksol-rhosol-Tempsol}
\beqa\label{eq:Rsol}
R(T)\,\Big|_\text{mod.\;FLRW}^\text{(rel-mat.\;sol.)} &=& 0\,,
\\[2mm]
\label{eq:Ksol}
K(T)\,\Big|_\text{mod.\;FLRW}^\text{(rel-mat.\;sol.)} &=& 
\frac{3}{2}\;\frac{1}{b^{4} + T^{4}}\,,
\eeqa
and also for the matter energy density $\rho$ 
and the matter temperature $\mathcal{T}$,
\beqa\label{eq:rhosol}
\rho(T)\,\Big|_\text{mod.\;FLRW}^\text{(rel-mat.\;sol.)}
&=& 
\rho_{0}\;\sqrt{\frac{b^{4}+T_{0}^{4}}{b^{4}+T^{4}}}\,,
\\[2mm]
\label{eq:Tempsol}
\mathcal{T}(T)\,\Big|_\text{mod.\;FLRW}^\text{(rel-mat.\;sol.)}
&=&
\mathcal{T}_{0}\;
\sqrt[8]{\frac{b^{4}+T_{0}^{4}}{b^{4}+T^{4}}}\,,
\eeqa
\esubeqs
for finite boundary conditions $\rho_{0} > 0$ and $\mathcal{T}_{0} > 0$
at $T=T_{0}>0$ [the actual value of $\rho_{0}$, for given $T_{0}$, 
follows from \eqref{eq:mod-Friedmann-equation-a}
and \eqref{eq:mod-Friedmann-T-odd-asol}].
The last result \eqref{eq:Tempsol} relies on the relation 
$\mathcal{T}(a) \propto \big[a^{4}\big]^{-1/4}$, 
as explained in the text below \eqref{eq:stand-FLRW-Temp-sol-with-t}.

The maximum value of the Kretschmann scalar \eqref{eq:Ksol}  
occurs at $T=0$ and is given by $K(0)= (3/2)\;b^{-4}$,
which allows for the interpretation of the parameter $b$ 
from the metric \emph{Ansatz} \eqref{eq:mod-FLRW-ds2}  
as the minimum curvature length scale of the resulting
spacetime manifold. The maximum
value of the matter density also occurs at $T=0$
and, from \eqref{eq:mod-Friedmann-equation-a}
and \eqref{eq:mod-Friedmann-T-odd-asol},
is given by $\rho(0) =(3/4)\, E^{2}_\text{planck}\,b^{-2}$,  
in terms of the reduced Planck energy
$E_\text{planck} \equiv \sqrt{1/(8\pi\, G_N)}   
\approx 2.44 \times 10^{18}\,\text{GeV}$.                
Similar results hold for a modified FLRW universe with nonrelativistic
matter, as discussed in 
Appendix~\ref{app:Modified-nonrelativistic-matter-dominated-FLRW-universe}. 
For completeness, we also present,
in Appendix~\ref{app:Modified-positive-CC-FLRW-universe}, 
a particular modified FLRW universe 
with a positive cosmological constant $\Lambda$. 


For the record, we mention that 
the $T$-even relativistic-matter solution $a(T)$
has the same eighth roots
as in \eqref{eq:mod-Friedmann-T-odd-asol} but now
with a plus sign before both roots. 
This $T$-even solution is perfectly smooth at $T=0$.
The results \eqref{eq:Rsol-Ksol-rhosol-Tempsol} hold also for
the $T$-even solution. 

We remark 
that it is possible to get the
$T$-even solution in a somewhat simpler form 
if we start from the definition 
$\widetilde{T}(\tau) \equiv \text{sgn}(\tau)\,\sqrt{\,\tau^{2} - b^{2}}\,$
(cf. Endnote~18 in Ref.~\cite{KlinkhamerSorba2014}), where  
the sign function is defined by $\text{sgn}(x)=x/\sqrt{x^2}$ for $x\ne 0$   
and $\text{sgn}(x)=0$ for $x = 0$. We then use the metric   
\beq\label{eq:mod-FLRW-ds2-Ttilde}
ds^{2}\,\Big|_\text{mod.\;FLRW}^{\left(\widetilde{T}-\text{coord.}\right)}
=
- \frac{\widetilde{T}^{\,2}}{b^{\,2}+\widetilde{T}^{\,2}}\;d\widetilde{T}^{\,2}
+ a^{2}\big(\tau\big)\;\delta_{kl}\,dx^k\,dx^l\,.  
\eeq
The corresponding modified Friedmann equation   
with relativistic matter now has the
following $\widetilde{T}$-even solution:        
\beq\label{eq:mod-Friedmann-T-even-asol-Ttilde}
a\big(\widetilde{T}\big)\,
\Big|_\text{mod.\;FLRW}^{\left(\widetilde{T}\text{-even\;rel-mat.\;sol.}\right)} =
\sqrt[4]{\Big(b^{\,2}+\widetilde{T}^{\,2}\Big)\Big/
         \Big(b^{\,2}+\widetilde{T}_{0}^{\,2}\Big)}\,.
\eeq
For the rest of the discussion, we again focus on the
$T$-odd solution \eqref{eq:mod-Friedmann-T-odd-asol}.

\section{Discussion}
\label{sec:Discussion}

In this article, we compare two spacetime manifolds.
The first spacetime manifold corresponds to the standard FLRW 
universe with metric \eqref{eq:stand-FLRW-ds2} for an extended cosmic time 
coordinate $T\in \mathbb{R}$ and the following cosmic scale factor solution 
for the case of relativistic matter:%
\beq\label{eq:singular-Friedmann-sol}
a(T)\,\Big|_\text{stand.\;FLRW}^\text{(rel-mat.\;sol.)}
=\text{sgn}(T)\, \sqrt[4]{T^{2}/T_{0}^{2}}\;,
\eeq
which extends the previous solution \eqref{eq:standard-Friedmann-asol}
for positive cosmic time coordinate $t$.
The line element of this first spacetime manifold is then given by
\beqa\label{eq:stand-FLRW-ds2-sol}
ds^{2}\,\Big|_\text{stand.\;FLRW}^\text{(rel-mat.\;sol.)}
&=&
- dT^{2} + \sqrt{T^{2}/T_{0}^{2}}\,\;\delta_{kl}\,dx^k\,dx^l\,,
\eeqa
with all spacetime coordinates $\{T,\, x^1,\, x^{2},\, x^{3}\}$
ranging over $\mathbb{R}$. 
The Kretschmann curvature scalar $K(T)$, 
the matter energy density $\rho(T)$, 
and the matter temperature $\mathcal{T}(T)$
obtained from \eqref{eq:singular-Friedmann-sol} 
are given by \eqref{eq:stand-FLRW-K-sol-with-t}, 
\eqref{eq:stand-FLRW-rho-sol-with-t}, 
and \eqref{eq:stand-FLRW-Temp-sol-with-t} 
with $t$ replaced by $T$, and they 
are seen to diverge as $T \to 0$.
The Einstein equation for the metric \eqref{eq:stand-FLRW-ds2-sol} 
is invalid at $T=0$ or, at least, ill defined.

With the definition $\eta \equiv \pm\,2\, \big[T^{2}\big]^{1/4}$,
the line element \eqref{eq:stand-FLRW-ds2-sol} becomes
conformally  flat, 
$ds^{2} = (1/4)\,\eta^{2} \big[d\eta^{2} + \delta_{kl}\,dx^k\,dx^l\big]$,
and corresponds to the background metric
used in Ref.~\cite{BoyleFinnTurok2018} with a slightly different notation.
In terms of the conformal  time $\eta$, 
the Kretschmann curvature scalar is given by
$K(\eta) \propto 1/\eta^{8}$, which diverges at $\eta=0$.

The second spacetime manifold corresponds to
a modified FLRW universe with metric \eqref{eq:mod-FLRW-ds2}
and  cosmic scale factor solution \eqref{eq:mod-Friedmann-T-odd-asol}.
The line element of this second spacetime manifold is given by
\beqa\label{eq:mod-FLRW-ds2-sol}
ds^{2}\,\Big|_\text{mod.\;FLRW}^\text{(rel-mat.\;sol.)}
&=&
- \frac{T^{6}}{\big(b^{4}+T^{4}\big)^{3/2}}\;dT^{2} 
+ \sqrt[4]{\frac{b^{4}+T^{4}}{b^{4}+T_{0}^{4}}}\;\delta_{kl}\,dx^k\,dx^l\,,
\eeqa
with all spacetime coordinates ranging over $\mathbb{R}$. 
The metric  \eqref{eq:mod-FLRW-ds2-sol} 
solves the Einstein equation with relativistic matter.
The corresponding quantities $K(T)$, $\rho(T)$, and  $\mathcal{T}(T)$
are given by \eqref{eq:Rsol-Ksol-rhosol-Tempsol}
and remain finite as $T \to 0$,
provided the length parameter $b$ is nonvanishing.

Both of the above metrics are degenerate,
$\det\,g_{\mu\nu} = 0$ at each spacetime point with $T=0$.
Observe, however, that the metric from \eqref{eq:mod-FLRW-ds2-sol} 
at each spacetime point with $T=0$
has a single vanishing eigenvalue (provided $b\ne 0$), 
whereas the metric from \eqref{eq:stand-FLRW-ds2-sol}
at the same spacetime point has three vanishing eigenvalues. 
Related to this last observation is the result that
the Kretsch\-mann curvature scalar $K$ 
is finite at $T=0$ for the metric \eqref{eq:mod-FLRW-ds2-sol}
with nonzero $b$
but is
singular at $T=0$ for the metric \eqref{eq:stand-FLRW-ds2-sol}
[these results can be seen explicitly in 
\eqref{eq:Ksol} for $b \ne 0$ and  $b = 0$, respectively].
Hence, the spacetime manifold \eqref{eq:mod-FLRW-ds2-sol},
with a nonzero length parameter $b$
and a particular differential structure,
may be considered to be a ``regularized'' version of
the spacetime manifold \eqref{eq:stand-FLRW-ds2-sol}.  
Corresponding results for a modified FLRW universe 
with a nonrelativistic matter component or a 
positive cosmological constant are given in 
Appendices~\ref{app:Modified-nonrelativistic-matter-dominated-FLRW-universe}
and \ref{app:Modified-positive-CC-FLRW-universe}.

We remark 
that, in general, a regularized theory may temporarily
lose certain desirable properties, which are only recovered as the
regulator is removed. An example is given by the lattice regularization
of flat-spacetime quantum field theory, where the full 
Poincar\'{e}-invariance group is recovered in the continuum
limit as the lattice spacing is taken to zero.
Our regularized Friedmann solution is also far from
perfect: the standard elementary-flatness condition breaks down 
at $T=0$, the location of the ``spacetime defect'' 
(cf. Appendix~D in Ref.~\cite{Klinkhamer2014-mpla}).
In addition, there is the $T=0$ discontinuity
in the $T$-odd cosmic scale factor  
solution \eqref{eq:mod-Friedmann-T-odd-asol}.
This discontinuity disappears in the corresponding
metric \eqref{eq:mod-FLRW-ds2-sol}, so that scalar, vector,
and tensor fields are unaffected by the $a(T)$ discontinuity 
at $T=0$. The spinor-field boundary conditions at $T=0^{+}$ 
and $T=0^{-}$ may require an appropriate $CP$ transformation.
In any case, the metric \eqref{eq:mod-FLRW-ds2-sol} provides 
a spacetime manifold without curvature
singularities, which allows for a meaningful study of the
behavior of relativistic matter in the very early universe.

The previously considered spacetime defect of          
Refs.~\cite{Klinkhamer2014-mpla,Klinkhamer2014-prd,KlinkhamerSorba2014,%
Guenther2017,KlinkhamerQueiruga2018,Klinkhamer2018-jpcs}
resulted from surgery on space. 
Here, we have considered surgery on cosmic time. 
But it is also possible to skip the surgery discussion.  
The new  metric  \textit{Ansatz} \eqref{eq:mod-FLRW-ds2}
simply replaces the FLRW \textit{Ansatz} \eqref{eq:stand-FLRW-ds2} and the 
resulting modified Friedmann equation \eqref{eq:mod-Friedmann-equation-a}
gives a universe without a
curvature singularity
but with a 3-dimensional spacetime defect,
which appears to be the ``lesser evil.''

It may be the case
that the length parameter $b$
entering the classical metric \eqref{eq:mod-FLRW-ds2-sol} is not just 
a mathematical artifact (``regulator'') but that it
traces back to the underlying theory of ``quantum spacetime.''
Even so, it is  unclear whether or not the length 
scale $b$ is determined by the Planck length    
$l_\text{planck} \equiv \sqrt{8\pi\,\hbar\, G_N/c^{3}} \approx
8.10 \times 10^{-35}\,\text{m}$, 
as the definitive 
quantum-spacetime theory has not yet been established.
See Ref.~\cite{Klinkhamer2007} for a general discussion
of a possible fundamental length scale that is
different from the Planck length.

Leaving aside a possible physical origin of the nonvanishing length parameter 
$b$ in the metric \eqref{eq:mod-FLRW-ds2-sol}, we observe that
the corresponding spacetime manifold is geodesically complete,
as long as the cosmic time coordinate $T$ has an extended
range, $T \in \mathbb{R}$. 
This manifold, then, has a pre-bounce phase for $T \leq 0$    
(a ``pre-big-bang'' phase, in standard terminology), 
which may or may not have produced relics in 
the present post-bounce universe for $T > 0$
(the present ``post-big-bang'' universe, in standard terminology).

In closing, we return to the condensed-matter-physics analogy mentioned
in the last sentence of Sec.~\ref{sec:Intro}.
In superfluid $^{3}$He experiments, there occur phase transitions 
between different topological phases~\cite{NissinenVolovik2018,Makinen-etal2019}. 
In one of these phases---the polar phase---the determinant of the 
effective tetrad field vanishes. This allows for a transition between 
two effective spacetimes with opposite chirality
if the system starts in the polar distorted 
A-phase, moves into the polar phase, and then returns back 
(cf. Fig.~1 in Ref.~\cite{Makinen-etal2019}).
The combined process in superfluid $^{3}$He 
(two phase transitions and the intermediate polar phase) 
is analogous to the spacetime defect of the 
regularized big bang singularity, which 
has a vanishing determinant of the spacetime metric.

In superfluid $^{3}$He, the system passes between the effective spacetimes 
via two subsequent phase transitions and an intermediate phase, 
as the temperature is changed by hand. 
In cosmology, the universe passes from one phase (pre-bounce)  
to another phase (post-bounce)  
via the spacetime defect, as the universe evolves 
forward by the reduced Einstein equations with appropriate initial conditions 
in the pre-bounce phase.

\begin{acknowledgments}

We thank G.E. Volovik for pointing
out Ref.~\cite{BoyleFinnTurok2018} and subsequent discussions,
and J.M. Queiruga and Z.L. Wang for useful comments 
on the manuscript.
Furthermore, we thank the referee for the suggestion to  
consider the case of a positive cosmological constant.

\end{acknowledgments}

\begin{appendix}
\section{Modified FLRW universe with nonrelativistic matter}
\label{app:Modified-nonrelativistic-matter-dominated-FLRW-universe}

In this appendix, we give some results 
for the  modified spatially flat FLRW universe 
with nonrelativistic matter
instead of the relativistic matter considered in
Sec.~\ref{sec:Modified-radiation-dominated-FLRW-universe}. 
Specifically, the equation of state \eqref{eq:mod-Friedmann-equation-c}
is replaced by
\beq
\label{eq:mod-Friedmann-equation-c-nonrel}
P(T)= 0\,,
\eeq
where $T$ is the cosmic time coordinate from \eqref{eq:T-def}.

The modified $T$-odd nonrelativistic-matter solution $a(T)$ 
from \eqref{eq:mod-Friedmann-equation-a},
\eqref{eq:mod-Friedmann-equation-b}, and
\eqref{eq:mod-Friedmann-equation-c-nonrel} is given by%
\beq\label{eq:mod-Friedmann-T-odd-asol-nonrel}
a(T)\,\Big|_\text{mod.\;FLRW}^\text{($T$-odd\;nonrel-mat.\;sol.)} = 
\begin{cases}
 + \sqrt[6]{\big(b^{4}+T^{4}\big)\big/\big(b^{4}+T_{0}^{4}\big)}\,,
 & \;\;\text{for}\;\; T > 0\,,
 \\[2mm]
 - \sqrt[6]{\big(b^{4}+T^{4}\big)\big/\big(b^{4}+T_{0}^{4}\big)}\,,
 &   \;\;\text{for}\;\; T \leq 0\,,
\end{cases} 
\eeq
with normalization $a(T_{0})=1$ for $T_{0}>0$.
The corresponding expressions for the Ricci curvature scalar $R$ and the 
Kretschmann curvature scalar $K$ are 
\bsubeqs\label{eq:Rsol-Ksol-nonrel}
\beqa
R(T)\,\Big|_\text{mod.\;FLRW}^\text{(nonrel-mat.\;sol.)} &=&
\frac{4}{3}\;\sqrt{\frac{1}{b^{4} + T^{4}}} \,,
\\[2mm]
K(T)\,\Big|_\text{mod.\;FLRW}^\text{(nonrel-mat.\;sol.)} &=& 
\frac{80}{27}\;\frac{1}{b^{4} + T^{4}}\,.
\eeqa
\esubeqs
Both curvature scalars are nonsingular at $T=0$ for $b\ne 0$  
and singular at $T=0$ for $b\to 0$.

Finally, the modified FLRW spacetime manifold with 
metric \eqref{eq:mod-FLRW-ds2}   
and cosmic scale factor solution \eqref{eq:mod-Friedmann-T-odd-asol-nonrel}
has the following line element:
\beqa\label{eq:mod-FLRW-ds2-sol-nonrel}
ds^{2}\,\Big|_\text{mod.\;FLRW}^\text{(nonrel-mat.\;sol.)}
&=&
- \frac{T^{6}}{\big(b^{4}+T^{4}\big)^{3/2}}\;dT^{2} 
+ \left[\,\frac{b^{4}+T^{4}}{b^{4}+T_{0}^{4}}\,\right]^{1/3}
\,\delta_{kl}\,dx^k\,dx^l\,,
\eeqa
with all spacetime coordinates ranging over $\mathbb{R}$.

For completeness, the $T$-even solution $a(T)$ has the same sixth roots
as in \eqref{eq:mod-Friedmann-T-odd-asol-nonrel} but now
with a plus sign before both roots. 
This $T$-even solution is perfectly smooth at $T=0$.
The results \eqref{eq:Rsol-Ksol-nonrel} and \eqref{eq:mod-FLRW-ds2-sol-nonrel} 
hold also for the $T$-even solution.
Again,  it is possible to get the
$T$-even solution in a somewhat simpler form if we
start from the definition 
$\widetilde{T}\equiv \text{sgn}(\tau)\, \sqrt{\,\tau^{2} - b^{2}}\,$.

\section{Modified FLRW universe with a positive cosmological constant}
\label{app:Modified-positive-CC-FLRW-universe}

In this appendix, we give some results 
for the  modified spatially flat FLRW universe
with a positive cosmological constant $\Lambda$,
which corresponds to a perfect fluid with constant vacuum energy density
$\rho_{V} =\Lambda$ and pressure $P_{V} =-\Lambda$.
The metric is again taken as \eqref{eq:mod-FLRW-ds2} for the
cosmic time coordinate $T \in \mathbb{R}$.

The dynamic equations are now given by 
\bsubeqs\label{eq:mod-Friedmann-equations-with-CC-abc} 
\beqa\label{eq:mod-Friedmann-equation-with-CC-a}
\left(1+ \frac{b^{4}}{T^{4}}\right)^{3/2} 
\left( \frac{1}{a(T)}\,\frac{d a(T)}{d T} \right)^{2}
&=& \frac{8\pi}{3}\,G_N\,\rho_{V}\,,
\\[2mm]
\label{eq:mod-Friedmann-equation-with-CC-b}
\frac{d}{d a} \bigg[ a^{3}\,\rho_{V}\bigg]+ 3\, a^{2}\,P_{V}&=&0\,,
\\[2mm]
\label{eq:mod-Friedmann-equation-with-CC-c}
P_{V} = -\rho_{V} =-\Lambda &<& 0\,,
\eeqa
\esubeqs
where \eqref{eq:mod-Friedmann-equation-with-CC-b} is satisfied automatically 
for the equation of state \eqref{eq:mod-Friedmann-equation-with-CC-c}.
With the following boundary conditions at $T_{0}>0$:
\bsubeqs\label{eq:mod-Friedmann-with-CC-asol-bcs}
\beqa
a(T_{0})&=&1 \,,
\\[2mm]
\big[(da/dT)/a\big]_{T=T_{0}}&>& 0\,,
\eeqa
\esubeqs
the $T$-even solution of \eqref{eq:mod-Friedmann-equation-with-CC-a} 
reads
\bsubeqs\label{eq:mod-Friedmann-with-CC-asol-HdS}
\beqa\label{eq:mod-Friedmann-with-CC-asol}
a(T)\,\Big|_\text{mod.\;FLRW}^\text{($T$-even\;CC\;sol.)} 
&=& 
\exp\left[\,
H_\text{dS}\,\left(\,\sqrt[4]{b^{4}+T^{4}_{\phantom{0}}}
-\sqrt[4]{b^{4}+T_{0}^{4}}\,\right)\,
\right]\,, 
\\[2mm]
\label{eq:mod-Friedmann-with-CC-HdS}
 H_\text{dS} &\equiv& \sqrt{8\pi\,G_N\,\Lambda/3}\,.
\eeqa
\esubeqs

We remark 
that the standard FLRW solution 
from a positive cosmological constant,
with $a(T) \propto \exp\big[H_\text{dS}\,T\big]$
for $T \in (-\infty,\,\infty)$,
has no big bang singularity at a finite value of $T$
and, hence, no need for
regularization.
Still, solution \eqref{eq:mod-Friedmann-with-CC-asol} 
may be of interest in that it joins an expanding
de-Sitter-type phase to a contracting de-Sitter-type phase, 
with a spacetime defect in between.
In fact, it is known that the complete
de Sitter spacetime~\cite{HawkingEllis1973} 
contains a patch with an expanding spatially flat FLRW
universe and another patch with a contracting spatially flat
FLRW universe and that the quantum fields in one 
patch   
may have unexpected interactions with those in the other 
patch
~\cite{KrotovPolyakov2010}.
These two FLRW patches of the standard de Sitter spacetime have
$a=0$ where they meet 
(the de Sitter manifold is, nevertheless, perfectly smooth everywhere), 
whereas the solution \eqref{eq:mod-Friedmann-with-CC-asol-HdS} 
has an exponentially small but nonzero value of $a(T)$ at $T=0$,
corresponding to the position of the spacetime defect.

\end{appendix}


\end{document}